\documentclass[11pt,oneside]{article}

\usepackage[utf8]{inputenc}
\usepackage[T1]{fontenc}
\usepackage{lmodern} 

\usepackage{graphicx}%
\usepackage{subcaption}
\usepackage{multirow}%
\usepackage{amsmath,amssymb,amsfonts}%
\usepackage{amsthm}%
\usepackage{mathrsfs}%
\usepackage{mathtools}
\usepackage[title]{appendix}%
\usepackage[svgnames,dvipsnames]{xcolor}%
\usepackage{textcomp}%
\usepackage{manyfoot}%
\usepackage{booktabs}%
\usepackage{algorithm}%
\usepackage{algorithmicx}%
\usepackage{algpseudocode}%
\usepackage{listings}%
\usepackage{braket}
\usepackage{bm}
\usepackage{here}
\usepackage{hhline}
\usepackage{authblk}
\usepackage{tabularx}
\usepackage{float}
\usepackage{caption}
\usepackage{tikz}
\usetikzlibrary{quantikz}
\usetikzlibrary{positioning,calc,decorations.pathreplacing}
\usetikzlibrary{arrows,chains,scopes}
\tikzset{
	ten/.style = {rectangle, draw=black},
	mat/.style = {ten, fill=blue!10, minimum size=1.6em},
	smat/.style={mat, minimum width=2.8em, node font=\small}, 
	fmat/.style ={ten, fill=blue!10, minimum width = 4.7em},
	node distance = 0.5em,
	uni/.style={circle, draw=black, minimum size=3em, fill=yellow!10},
	bguni/.style={uni, minimum size=3.9em},
	node font=\scriptsize
}
\usepackage[colorlinks,citecolor=DarkGreen,linkcolor=FireBrick,linktocpage,unicode]{hyperref}

\newcommand*{\figref}[1]{\figurename~\ref{#1}}
\newcommand*{\tblref}[1]{\tablename~\ref{#1}}

\title{\textbf{Time series generation for option pricing on quantum computers using tensor network}}

\author{Nozomu Kobayashi}

\author{Yoshiyuki Suimon}

\author[2]{Koichi Miyamoto}

\affil[1]{Data Science Department, Nomura Securities Co., Ltd., Tokyo, Japan}

\affil[2]{Center for Quantum Information and Quantum Biology, The University of Osaka, Osaka, Japan}

\begin{document}

\maketitle

\begin{abstract}
Finance, especially option pricing, is a promising industrial field that might benefit from quantum computing.
While quantum algorithms for option pricing have been proposed, it is desired to devise more efficient implementations of costly operations in the algorithms, one of which is preparing a quantum state that encodes a probability distribution of the underlying asset price.
In particular, in pricing a path-dependent option, we need to generate a state encoding a joint distribution of the underlying asset price at multiple time points, which is more demanding.
To address these issues, we propose a novel approach that uses a Matrix Product State (MPS), which can be encoded into a state of qubits, as a generative model for time series generation.
We focus on the training of such an MPS and present its procedure in detail.
To validate our approach, taking the Heston model as a target, we conduct numerical experiments to generate time series in the model.
Our findings demonstrate the capability of the MPS model to generate paths in the Heston model, highlighting its potential for path-dependent option pricing on quantum computers.
\end{abstract}

\section{Introduction} \label{sec:introduction}
Quantum computing has recently received growing attention in industries as it is expected to achieve significant speed-up for many computational problems.
One of the fields people expect to benefit from quantum computers is finance, where several potential applications have already been considered, including portfolio optimization, risk management, and option pricing. 
See \cite{orus2019quantum,bouland2020prospects,egger2020quantum,herman2022survey,herman2023quantum} for comprehensive reviews of these topics.
The application we would like to address in this paper is option pricing with the Monte Carlo method, which is theoretically shown in \cite{montanaro2015quantum} that it can achieve quadratic speed up compared with the classical counterpart.
An {\it option}, a type of derivative, is a financial contract that gives the buyer the right to buy or sell an underlying asset, such as a stock, an interest rate index, or a foreign exchange rate, at a predetermined price and date.
However, we use the term ``option'' to refer to a wider range of contracts in which the buyer receives an amount of money ({\it payoff}) linked to an underlying asset, since such contracts, including some examples presented later, are often called options.
Apart from underlying asset markets, options offer investors alternative opportunities for profits and hedging various risks.
Based on the various needs of investors, a wide variety of options have been introduced.
Therefore, it is a central problem for financial institutions to price options in a suitable manner in order to trade and manage them.

Following the quantum algorithm for Monte Carlo integration (QMCI) in \cite{montanaro2015quantum}, which is based on Quantum Amplitude Estimation (QAE) \cite{brassard2002quantum}, quantum approaches for Monte Carlo-based option pricing have been proposed \cite{rebentrost2018quantum,stamatopoulos2020option,kaneko2022quantum}. 
The option price (or option premium) is expressed as the expectation of the payoff in a stochastic model of the time evolution of the underlying asset price, and thus its evaluation can be sped up by QMCI.
Although these sound promising, it is unclear whether they will be practicable on real quantum devices, as among the procedures are costly operations such as preparing quantum states that encode probability distributions of the underlying asset price in the amplitude.
Using the famous Grover-Rudolph method \cite{grover2002creating}, it is possible to generate such a state if we can express a cumulative distribution in a given interval by a simple formula \cite{kaneko2022quantum}.
However, this approach requires many runs of quantum circuits for arithmetic operations \cite{MunozCores2022}, which results in large gate costs.
Moreover, the task becomes more demanding, if we would like to price not only European-type options, in which the payoff depends on the underlying asset price on a single date, but also path-dependent ones, whose payoff depends on the asset price at multiple dates.
That is, we need to generate a quantum state that encodes the joint distribution of the asset price at the different time points.
In other words, we need to prepare a state encoding the distribution of time series, or {\it paths}, of the asset price.

To reduce the cost of state preparation, some alternative methods using no arithmetic circuit have been proposed.
Along with methods with rigorous error bounds \cite{Sanders2019,wang2021fast,Bausch2022fastblackboxquantum,Wang_2022,mcardle2022quantum,rattew2022preparing,Sanchez2023,Moosa_2023}, there are some heuristic methods, which might be able to reduce the cost further \cite{zoufal2019quantum,GarciaRipoll2021quantuminspired,Endo2020,Holmes2020,Sanchez2023}.
In particular, some of such methods use quantum generative modeling.
\cite{zoufal2019quantum} proposed the quantum generative adversarial networks (GAN), claiming it can reduce the number of quantum gates for loading probability distributions.
The successive work \cite{fuchs2023hybrid} generalized the framework of quantum GAN by leveraging the Wasserstein loss function, which is in line with the development of the classical Wasserstein GAN.
However, as far as the authors know, there is no previous study on preparing a state encoding the time series distribution via generative modeling in the context of option pricing on a quantum computer.
That is what we would like to address in this paper.

To tackle this problem, we make good use of Matrix Product State (MPS) as a generative model.
MPS, a kind of Tensor Network (TN), which was originally developed to represent quantum many-body wave functions efficiently on classical computers, is a powerful and flexible tool to approximate higher-order tensors as the network of lower-order tensors, thereby reducing the complexity of original tensors.
This capability has led to recent applications of MPS in machine learning \cite{novikov2016exponential,stoudenmire2016supervised,huggins2019towards}, which includes the particular focus on generative modeling \cite{han2018unsupervised,cheng2019tree,liu2023tensor}.
We aim to utilize the MPS-based generative modeling for time series.
Although generative modeling via MPS is classically tractable, it is strongly related to quantum computing since, given an MPS, we can generate the state on qubits with the corresponding wave function by a quantum circuit in an efficient manner \cite{ran2020encoding,rudolph2022synergy,rudolph2023decomposition}.
In fact, there are some studies on MPS-based methods for function-encoding quantum state preparation \cite{GarciaRipoll2021quantuminspired,Holmes2020}, although they are not on a time series distribution.
Then, we conceive the following approach for option pricing on a quantum computer with the aid of MPS: we can find the MPS that generates asset price paths in a given model by a classical computer, and then use the corresponding state generation circuit in QMCI.

That being said, in this paper, we investigate the MPS model for generating time series with the goal of pricing options in finance.
Specifically, we build a generative model where an MPS serves as an ansatz and is trained to minimize the Kullback-Leibler (KL) divergence between it and the desired distribution of asset price paths.
As an illustration of our methodology, we focus on the Heston model \cite{heston1993closed}, a well-known and widely used stochastic process in the financial industry.
To validate our approach, we generate price paths using the MPS model and compute prices of path-dependent options through classical Monte Carlo simulations.
We find that the MPS model can successfully generate the price paths of the Heston model and be used in pricing these options.

The remainder of this paper is organized as follows.
In Section \ref{sec:financial_backgrounds}, we review the basics of option pricing and introduce the Heston model. We also briefly comment on the framework of the quantum algorithm for option pricing.
Section \ref{sec:MPS} is devoted to proposing MPS for generative modeling of time series.
In Section \ref{sec:experiment}, we report the validity of the proposed MPS model in option pricing by conducting numerical experiments.
Finally, Section \ref{sec:conclusion} concludes this paper with a summary and outlook.

\section{Financial background}\label{sec:financial_backgrounds}
In this section, we aim to provide a review of option pricing with a particular focus on the Monte Carlo method.
First, we will outline the concept of options, and explain how to price them.
Since we consider the Heston model in this paper, we briefly introduce that model.
Finally, we comment on option pricing with a classical or quantum computer.

\subsection{Options}\label{sec:option}
An option is originally a financial contract between two parties, one of which, the option holder, possesses the right, but not the obligation, to either buy or sell an underlying asset at a pre-determined price ({\it strike}), on a future date.
If the holder has a right to buy (resp. sell) the underlying asset, the option is referred to as a call (resp. put) option.
Let $S_t$ be an asset price at time $t$ and consider a call option written on that asset exercisable on a maturity date $t=T$. 
At the maturity date, the holder exercises a right to buy the asset if its price is higher than the strike $K$.
This is equivalent to getting a profit $S_T - K$, since the asset is worth $S_T$.
Conversely, if $S_T<K$, the holder chooses not to exercise the right and receives nothing. 
In total, this option is equivalent to the contract in which the option holder receives the payoff
\begin{align}
    f_{\rm pay} (S_T) = \max\{S_T - K,0\} \, ,
\end{align}
at $t=T$.

Derived from this kind of option called a {\it European option}, many kinds of options with various payoffs have been developed.
In particular, our study focuses on path-dependent options, whose payoff depends not on the asset price at a single time point but on the values at multiple time points or the entire path of the asset price. 
In the following, we present examples of such options considered in this work.

\paragraph*{Asian option} An Asian option has a payoff depending on the average price of the underlying asset.
Explicitly, a payoff of an Asian call option with a strike $K$ is given by 
\begin{align}
    f_{\rm pay}(\{S_t\}_{0 \leq t \leq T}) = \max \left\{ \frac{1}{T}\int_0^T S_t dt - K,0 \right\}.
\end{align}

\paragraph*{Lookback option} A lookback option has a payoff depending on the maximum or minimum price of the underlying asset over the life of the option.
A payoff of a Lookback call option with a strike $K$ is given by 
\begin{align}
    f_{\rm pay}(\{S_t\}_{0 \leq t \leq T}) = \max \left\{ \max_{0 \leq t \leq T} S_t - K , 0 \right\}.
\end{align}

\paragraph*{Barrier option} 
A barrier option has a payoff which depends on whether an underlying asset price touches the barrier level $B$ before the maturity date.
While there are various types of barrier options, we consider an up-and-out barrier call option with a strike $K$, whose payoff is given by 
\begin{align}
    f_{\rm pay}(\{S_t\}_{0 \leq t \leq T}) =
    \begin{cases}
    \max \left\{S_T - K,0  \right\}& (\max_{0 \leq t \leq T} S_t  \,  < B  )\\
        0 & (\mathrm{otherwise})
    \end{cases}.
\end{align}
That means that the up-and-out call option expires worthless when $S_t$ crosses the barrier level $B$.

\subsection{Models}

In order to trade these financial products in the market, it is essential to determine their theoretical fair prices.
The price of an option at the present time $t=0$ is given by the expectation value (see textbooks \cite{hull1993options,shreve2005stochastic,shreve2004stochastic} for the details),
\begin{align}
    V_0 = E[f_{\rm pay}(\{S_t\}_{0 \leq t \leq T})] \, ,
    \label{eq:OpPrice}
\end{align}
in the risk-neutral measure. Note that, for the sake of simplicity, we assume the risk-free rate to be 0 here and hereafter.

As such, the central problem in option pricing has revolved around the calculation of the above expectation value in a stochastic model of $S_t$.
The celebrated Black-Scholes (BS) model \cite{black1973pricing,Merton1973} assumes that, in the risk-neutral measure, the dynamics of $S_t$ is given by the following stochastic differential equation (SDE) $dS_t = \sigma dW_t$, where $W_t$ denotes a Wiener process and the volatility $\sigma$ is a positive constant.

Besides the BS model, many kinds of advanced models have been considered so far.
In this paper, we focus on the Heston model \cite{heston1993closed}, which is widely used in the financial industry.
It is a kind of stochastic volatility (SV) model with the SDE in the following form,
\begin{align}
    dS_t = \sigma_t dW_t,
    \label{eq:SDE}
\end{align}
where the volatility $\sigma_t$ is not a constant but some stochastic process.
Leaving further details to Appendix \ref{sec:Heston}, we note that the Heston model has four parameters $\kappa, \theta, \xi$, and $\rho$, and thus provides more flexibility in fitting model predictions for option prices to actual market prices, compared to the BS model with its single parameter $\sigma$.

\subsection{Option pricing with the Monte Carlo method on classical and quantum computers}

For the BS model and simple products such as European options, analytical pricing formulas are available~\cite{hull1993options}.
On the other hand, for advanced models such as the Heston model and complicated products listed above, analytical formulas are unavailable, and thus we resort to numerical methods.
Among them, we hereafter focus on the Monte Carlo method.
This requires generating numerous price paths for the underlying asset $S_t$ based on the assumed model like Eq.\eqref{eq:SDE}. 
In reality, we cannot generate paths $\{S_t\}_t$ for continuous time $t$ but discretized paths $\{S_{t_j}\}_{j=0,1,...,M}$ consisting of a finite number of time points $t_0=0<t_1<\cdots<t_M=T$.
Following the generation of $N$ simulated price paths $\{S_{t_j}^{i}\}_{i,j}$, where $i$ is the index of the sample paths, the price of the option is estimated by 
\begin{align}
    V_0 \simeq \frac{1}{N}\sum_{i=1}^N f_{\rm pay}(\{S_{t_j}^{i}\}),
\end{align}
where the payoff function $f_{\rm pay}$ is somehow approximated if it is defined with a path on continuous time.
This approach allows us to numerically evaluate complex options with path dependence.
However, it should be noted that the Monte Carlo method can be computationally intensive particularly when generating a large number of paths is required: for accuracy $\epsilon$ in $V_0$, the number of paths $N$ scales as $O(1/\epsilon^2)$.

This motivates us to consider applying QMCI.
For accuracy $\epsilon$, its complexity is $O(1/\epsilon)$, which means the quadratic speed-up of the Monte Carlo method.
Leaving more details to Appendix \ref{sec:QM}, we note that loading the probability distribution of paths into a quantum state is one of the nontrivial steps in QMCI-based option pricing.
That is, we need to generate the following quantum state
\begin{align}
    \sum_{\{S_{t_j}\}_j} \sqrt{p(\{S_{t_j}\}_j)} \ket{\{S_{t_j}\}_j} \, ,
    \label{eq:TargetState}
\end{align}
where $p(\{S_{t_j}\}_j)$ is a probability associated with the path $\{S_{t_j}\}_j = (S_{t_0}, S_{t_1}, \cdots S_{t_M})$ and $\ket{\{S_{t_j}\}_j}=\ket{S_{t_0}}\cdots\ket{S_{t_M}}$ is the computational basis state that represents the finite-precision binary representations of $S_{t_0},\cdots,S_{t_M}$.
However, as explained in Section \ref{sec:introduction}, implementing this state preparation may be computationally demanding, and it is desired to find an efficient way to realize it. 

\section{Generative model using MPS}\label{sec:MPS}
In this section, we elaborate on our proposal for leveraging MPS for a time-series generative model.
Let $x_t=(x_1,\cdots,x_M)$ be a sequence of random variables driven by a stochastic process and $x_t^i$ be its $i$th realization.
By collecting $x_t^i$, we can construct a dataset of time series with $N$ samples, $\mathcal{T} = \{x_t^i\}_{i=1}^N$.
The objective of a generative model is to construct a probabilistic model $T_\theta$, which can generate synthetic data $\hat{x}_t  \sim p_\theta (x_t)$ so that $p_\theta (x_t)$ is as close as possible to the true probability distribution of $x_t$.

\subsection{MPS for time series generation}
In this study we propose employing MPS as a generative model $T_\theta$. 
To this end, we first need to discretize our dataset so that
\begin{align}
    \mathcal{T} \rightarrow \bar{\mathcal{T}} = \{ \bar{x}_t^i \}_{i=1}^N, 
\end{align}
Here, we transform continuous variables $x_t$ into integers in $\{0,\cdots,2^m-1\}$ corresponding to $m$-bit binary values as
\begin{align} \label{eq:discretize_price}
    x_j \rightarrow  \bar{x}_j = \left\lfloor ( 2^m - 1) \times \frac{x_j - x_{\rm min}}{x_{\rm max} - x_{\rm min}} \right\rfloor
\end{align}
with $x_{\rm min}:=\min_{i,j} x_j^i$ and $x_{\rm max}:=\max_{i,j} x_j^i$.
This corresponds to the discretization of each variable $x_j$ by the $2^m$ grid points, among which the one labeled by $\bar{x}_j\in\{0,\cdots,2^m-1\}$ is given by
\begin{align}
    x_{j} = x_{\rm min} + \frac{\bar{x}_j}{2^m-1}(x_{\rm max}-x_{\rm min}).
    \label{eq:IntToReal}
\end{align}
This can naturally be fit in the computational basis of a quantum state.
Then, the MPS $\Psi$ is introduced as a rank-$M$ tensor with each leg having dimension $2^m$.
Each entry in $\Psi$ is indexed by $\bar{x}_t=(\bar{x}_1,\ldots,\bar{x}_M)$, which can take $(2^m)^M$ values, and the entry associated with fixed $\bar x_t$ is a real number
\begin{align}
    \Psi(\bar x_t) = \sum_{\alpha_1=0}^{D_1-1}\cdots\sum_{\alpha_M=0}^{D_M-1} A^{(1)}_{\bar x_1 \alpha_1} A^{(2)}_{\alpha_1 \bar x_2 \alpha_2} \cdots A^{(M)}_{\alpha_{M-1} \bar x_M} \, .  \label{eq:our_MPS}
\end{align}
Here, $A^{(j)}_{\alpha_{j-1} \bar x_j \alpha_j} \in \mathbb{R}^{D_{i-1} \times 2^m \times D_i}$ is a rank-3 tensor, and $D_i \in \mathbb{N}_{> 0}$ is referred to as bond dimension with $D_0 = D_M = 0$.
The $j$-th entry $\bar x_j\in\{0,\cdots,2^m-1\}$ in $\bar{x}_t$ corresponds to the second leg of the $j$-th tensor, whose dimension $2^m$ are called physical dimension.
Having the MPS $\Psi$, we calculate the probability distributions of $\bar x_t$ by 
\begin{align}
    p_\theta (\bar x_t) = \frac{ |\Psi (\bar x_t)|^2}{Z} , \, 
\end{align}
with the partition function $Z = \sum_{\bar x_t} \left|\Psi (\bar x_t)\right|^2$.
For an MPS, $Z$ is given by its self-contraction,
\begin{align}
    Z= \sum_{\bar{x}_1,\cdots,\bar{x}_M}\sum_{\alpha_1,\cdots,\alpha_{M}}\sum_{\beta_1,\cdots,\beta_{M}} A^{(1)}_{\bar x_1 \alpha_1} A^{(2)}_{\alpha_1 \bar x_2 \alpha_2} \cdots A^{(M)}_{\alpha_{M-1} \bar x_M}
    A^{(1)}_{\bar x_1 \beta_1} A^{(2)}_{\beta_1 \bar x_2 \beta_2} \cdots A^{(M)}_{\beta_{M-1} \bar x_M}.
\end{align}
This can be calculated in $O(2^m M \overline{D}^3)$ time, where $\overline{D}=\max\{D_1,\ldots,D_{M-1}\}$, as explained in reviews on tensor network (see, e.g., Sec. 5.1.1 (5) in \cite{ORUS2014117}).

It is worth noting that previous studies \cite{han2018unsupervised,cheng2019tree,liu2023tensor} on generative modeling with MPS typically assign each digit of $\bar x$ into a different tensor, which means that, in our time series setting, each $\bar{x}_j$ is further decomposed into $\bar{x}_j = a_{j,0} + a_{j,1} \times 2 + \cdots a_{j,m-1} \times 2^{m-1} $ with $a_{j,k} \in [0,1]$.
Consequently, the model would provide us with the following structure 
\begin{align}
    \Psi(\bar x_t) = \sum_{\tilde\alpha_1,\cdots,\tilde\alpha_{Mm}} \tilde{A}^{(1)}_{a_{1,0}\tilde\alpha_1 } \tilde{A}^{(2)}_{\tilde\alpha_1 a_{1,1} \tilde\alpha_2} \cdots A^{(Mm)}_{\tilde\alpha_{Mm-1}a_{M,m-1}} \, .
\end{align} 
In such a formulation, the MPS model deals with correlation between $\bar x_{t-1}$ and $\bar x_t$ only through the first and last digits of them, which we suppose may result in poor expressive power for time series.
This consideration motivates us to formulate our model as in Eq.\,\eqref{eq:our_MPS}.

\subsection{Training MPS model}
In order to train our MPS model, we introduce the KL divergence as a measure to discern and evaluate the dissimilarity between probability distributions,
\begin{align}
    D_{KL} \left( p_\theta (\bar x_t) | \pi (\bar x_t) \right) = \sum_{i=1}^N p_\theta (\bar x_t^i) \ln \left(\frac{p_\theta (\bar x_t^i)}{\pi (\bar x_t^i)} \right) \, . \label{eq:KL_divergence}
\end{align}
Here, we denote the probability distribution of the discretized random variable $\bar x_t$ as $\pi (\bar x_t)$, which is a pre-specified distribution we aim to reproduce. 
Given that the KL divergence represents the distance between two distributions, our primary aim is to minimize Eq.\,\eqref{eq:KL_divergence}.
Minimizing the KL divergence can be achieved by minimizing the following negative log-likelihood,
\begin{align}
    \mathcal{L} = - \frac{1}{|\mathcal{T}|} \sum_{\bar{x}_t^i} \ln p_\theta (\bar x_t^i) \, , 
\end{align}
which is equivalent to the KL divergence up to a constant.
Thus, we adopt the negative log-likelihood as the objective function for training the MPS model.

The training of tensor components in the MPS is conducted through the gradient descent technique.
While in usual manner of the gradient descent whole tensor elements are simultaneously updated, in this work we take advantage of the sweeping algorithm as in \cite{han2018unsupervised}.
In this algorithm, each of adjacent tensor component pairs is iteratively updated while the others unchanged, and the bond dimension between the components is automatically adjusted, which is an advantage compared with the usual gradient descent approach.
This is in fact similar to the density matrix renormalization group (DMRG) algorithm.
For the full details of this update procedure, see Appendix \ref{sec:MPSUpdate}.

\subsection{Generating samples from the MPS \label{sec:sampling}}
While our final goal is to prepare a quantum state encoding a probability distribution, one can generate samples from the MPS model.
Thanks to the fact that the partition function of MPS can be calculated efficiently, it enables us to draw samples directly without help of such as Gibbs sampling.
To begin with, we take the leftmost tensor in MPS and compute the following marginal probability:
\begin{align}
    P(\bar x_1) = \frac{\sum_{\bar{x}_2,\cdots,\bar{x}_M = 0}^{2^m-1} |\Psi(\bar x_1,\bar{x}_2,\cdots,\bar{x}_M)|^2}{Z} \, ,
\end{align}
where we only contract the second and below physical legs of MPS in the numerators.
According to $P(\bar x_1)$, we can draw a sample of $\bar x_1$. 
Given that sample, the drawn probability of the next one $\bar x_2$ is given as the conditional probability:
\begin{align}
    P(\bar x_2 | \bar x_1) = \frac{P(\bar x_2, \bar x_1)}{P(\bar x_1)}  \, , 
\end{align}
where $P(\bar x_2, \bar x_1)$ can similarly be calculated as $P(\bar x_1)$.
Using this conditional probability $P(\bar x_2 | \bar x_1) $, $\bar x_2$ can be drawn.
Repeating this procedure one-by-one, we are able to obtain the series of samples ${\bar x_t} = (\bar x_1, \bar x_2 ,\cdots , \bar x_n)$.
Then, each integer $\bar{x}_j$ is converted to the real value $x_j$ by Eq.~\eqref{eq:IntToReal}, which can be regarded as an approximation of the original random variable with rounding error of order $O(2^{-m})$.

The time complexity of sampling one value of $\bar{x}_t$ is $O(2^m M \overline{D}^3)$.
Here, the dominant contribution comes from calculating the (conditional) probability mass functions (PMFs) $P(\bar x_1)$, $P(\bar x_2 | \bar x_1)$, and so on, by partial self-contraction of $\Psi$, whose computational time is bounded by that of full self-contraction and thus of order $O(2^m M \overline{D}^3)$.
Once we get the PMF of each $\bar{x}_j$, which consists of $2^m$ real numbers, sampling from it takes $O(2^m)$ time~\cite{fishman2013monte}.
To get one sample of $\bar{x}_t$ thus takes $O(2^mM)$ time, which is subdominant compared to obtaining the PMFs.

\subsection{Quantum circuit to encode the MPS \label{sec:circuit}}

\begin{figure}[t]
\centering
\begin{subfigure}{1\textwidth}
    \centering
\begin{tikzpicture}
    \node[draw, shape=circle,minimum width=3.5em] (v0) at (0,0) {$A^{(1)}$};
    \node[draw, shape=circle,minimum width=3.5em] (v1) at (1.5,0) {$A^{(2)}$};
    \node[] (v2) at (3,0) {$\cdots$};
    \node[] (v3) at (0.5,-1) {};
    \node[draw, shape=circle,minimum width=3.5em] (v4) at (4.8,0) {$A^{(M)}$};
    \draw [thick]
    (v0) -- (v1) node[midway, above] {$\alpha_1$}
    (v0) -- (v0 |-  v3.north) node[midway, right] {$\bar{x}_1$}
    (v1) -- (v1 |- v3.north) node[midway, right] {$\bar{x}_2$}
    (v2) -- (v4) node[midway, above] {$\alpha_{M-1}$}
    (v4) -- (v4 |- v3.north) node[midway, right] {$\bar{x}_M$}
    (v1) -- (v2) node[midway, above] {$\alpha_2$};
\end{tikzpicture}
    \caption{}
    \label{fig:MPS}
    \end{subfigure}

    \vspace{1em}

\begin{subfigure}{1\textwidth}
    \centering
    \begin{quantikz}
& \lstick{$\ket{0}$} & \qwbundle{d_1} & \qw & \qw & \qw & \gate[2]{{\Large W^{(1)}}} & \qw \arrow[r, draw=none, "\bar{x}_1", xshift=-10ex, yshift=0.5ex]  & \qw & \cdots & & \qw & \qw& \qw& \qw  & \\
& \lstick{$\ket{0}$}  & \qwbundle{d_2} & \qw & \qw & \qw & \qw\arrow[r, draw=none, "\alpha_1", xshift=0.5ex, yshift=0.5ex] & \gate[2]{W^{(2)}} & \qw \arrow[r, draw=none, "\bar{x}_2", xshift=-4.2ex, yshift=0.5ex] & \cdots & & \qw & \qw& \qw& \qw &\\
& \lstick{$\ket{0}$} & \qwbundle{d_3} & \qw & \qw & \qw & \qw & & \qw \arrow[r, draw=none, "\alpha_2", xshift=-4ex, yshift=0.5ex] & \cdots & & \qw & \qw& \qw& \qw & \\
& & & & & & & \vdots & & & & & & & \\
& \lstick{$\ket{0}$} & \qwbundle{d_{M-1}} & \qw &  \qw & \qw & \qw & \qw & \qw & \cdots & & \gate[2]{W^{(M-1)}} & \qw \arrow[r, draw=none, "\bar{x}_{M-1}", xshift=-4.5ex, yshift=0.5ex] & \qw & \qw &\\
& \lstick{$\ket{0}$} & \qwbundle{d_M} & \qw & \qw & \qw & \qw & \qw & \qw & \cdots & & \qw\arrow[r, draw=none, "\alpha_{M-1}", xshift=2ex, yshift=0.5ex]& \qw  & \gate[1]{W^{(M)}} & \qw \arrow[r, draw=none, "\bar{x}_M", xshift=-3.5ex, yshift=0.5ex] &
\end{quantikz}

    \caption{}
    \label{fig:MPSCircuit}
    \end{subfigure}
    \caption{(a) Tensor network diagram that represents the MPS $\Psi$. (b) Quantum circuit to generate the MPS-encoding state. The symbol such as $\bar{x}_j$ and $\alpha_j$ above the wire represents the corresponding physical or bond index.}
    \label{fig:MPSToCircuit}
\end{figure}

Given an MPS $\Psi$, we can generate the quantum state that encodes $\Psi$ in the amplitudes:
\begin{align}
    \ket{\Psi} \coloneqq \frac{1}{\sqrt{Z}}\sum_{\bar{x}_1,\cdots,\bar{x}_M = 0}^{2^m-1} \Psi(\bar x_1,\cdots,\bar{x}_M)\ket{\bar x_1}\cdots\ket{\bar x_M}.
\end{align}
This is a state on a system with $M$ registers, each of which consists of at least $m$ qubits, and $\ket{\bar{x}_j}$ is the computational basis state on the $j$-th register that corresponds to the binary representation of the integer $\bar{x}_j$.
As considered in \cite{ran2020encoding,miyamoto2023extracting}, generating this state is done by the quantum circuit shown in \figref{fig:MPSToCircuit}.
The $j$-th register has $d_j$ qubits, where
\begin{align}
    d_j \coloneqq
    \begin{cases}
        m & ; \ j=1 \\
        \max\{\lceil \log_2 D_{j-1} \rceil, m \} & ; \ \text{otherwise}
    \end{cases}
    .
\end{align}
$W^{(j)}$ is a $2^{d_j+d_{j+1}} \times 2^{d_j+d_{j+1}}$ unitary, except $W^{(M)}$ is $2^{d_M} \times 2^{d_M}$.
$W^{(j)}$ encodes the tensor $A^{j}$ in its entries as\footnote{Precisely speaking, the MPS needs to be transformed into the right canonical form before encoding $A^{(j)}$ into $W^{(j)}$ \cite{miyamoto2023extracting}. This is not necessary for sampling from the MPS described in Sec. \ref{sec:sampling}, and thus we do not do it in this paper.}
\begin{align}
    & \bra{\bar{x}_1}\bra{\alpha_1}W^{(1)}\ket{0}\ket{0} = (A^{(1)})_{\bar{x}_1 \alpha_1}, \nonumber \\
    & \bra{\bar{x}_j}\bra{\alpha_j}W^{(j)}\ket{\alpha_{j-1}}\ket{0} = (A^{(j)})_{\alpha_{j-1} \bar{x}_j \alpha_j}  \quad \text{for } j=2,\ldots,M-1,  \nonumber \\
    & \bra{\alpha_{M-1}}W^{(M)}\ket{\bar{x}_M} = (A^{(M)})_{\alpha_{M-1}\bar{x}_M}.
\end{align}
Any $d$-qubit unitary can be implemented as a quantum gate with $O(4^n)$ elementary gates~\cite{Vartiainen2004}, and so is $W^{(j)}$ with $O(4^{d_j+d_{j+1}})$ gates.
Thus, the quantum circuit in \figref{fig:MPSToCircuit} is constructed with
\begin{align}
O\left(\sum_{j=1}^{M-1} 4^{d_j+d_{j+1}}+4^{d_M}\right)=O\left(M\left(\max\{\overline{D},2^m\}\right)^4\right)
\label{eq:gatecost}
\end{align}
elementary gates.
Compared with the $O(2^{mM})$ gate cost for generating a general $mM$-qubit state~\cite{Mottonen2005}, which is exponential in $M$, the cost \eqref{eq:gatecost} is polynomial in $M$.

\section{Numerical experiments}\label{sec:experiment}
In this section, we report numerical experiments to validate our proposed model.
To this end, we train the MPS model to generate asset price paths in the Heston model, using the paths generated classically. 
We leverage the trained MPS model to generate time series samples, by means of which we compute prices of various path-dependent options with the Monte Carlo method.
In order to evaluate its performance, we compare the results with those obtained by the Heston model itself.

\subsection{Setting}
We here summarize the setting of our numerical experiments.
We set these parameters as $\kappa = 1.0$, $\theta = 0.04$, $\xi = 2$, and $\rho = - 0.7$.
The initial values of the asset price $S_0$ and the squared volatility $v_0$ are set to 100 and 0.04, respectively.
To generate paths in the Heston model, we adopt the simple Euler-Maruyama discretization scheme \cite{Maruyama1955}, that is, we discretize the original stochastic process as below: 
\begin{align}
S_{i+1} &= S_i + \sqrt{\nu_i} S_i \sqrt{\Delta t} Z^S  \, , \nonumber \\ 
\nu_{i+1} &= V_i + \kappa(\theta - \nu_i) \Delta t + \xi \sqrt{\nu_i} \sqrt{\Delta t} Z^\nu \, ,
\label{eq:EulerMaruyama}
\end{align}
where $\Delta t = t_{i+1} - t_{i}$ is a time step, and $Z^S$ and $Z^\nu$ are random variables drawn from the bivariate standard normal distribution with correlation $\rho$. To avoid a negative variance, we assume the reflection positivity, $\nu_i = - \nu_i$ if $\nu_i < 0$.
In the experiment, we set the time step as $\Delta t = 1/ 250$, corresponding to daily observations, and the length of time series as $M=5$.
With these settings, we then generate $N=10000$ paths of the Heston model.

In the training of the MPS model, the physical dimensions are varied as $2^m$ with $m=4,5,6$. 
That means we discretize each price into $m$-bit binary values.
Concretely, we adopt the aforementioned way,
\begin{align}
    S_t \rightarrow  \bar{S_t} = \left\lfloor ( 2^m - 1) \times \frac{S_t - S_{\rm min}}{S_{\rm max} - S_{\rm min}} \right\rfloor
\end{align}
with $S_{\rm min}$ and $S_{\rm min}$ are the minimum and maximum of the asset price, respectively, over sample paths and $n$ time points.
Since our training procedure allows us to optimize bond dimensions, we fix the maximum value of bond dimensions $D_{\rm max}$ in the training. 
In our numerical experiment, we consider different values of $D_{\rm max} = 64,100,150$.

After the training, we generate $N=10000$ paths using the resultant model, converting the model output integers to real values as Eq.\,\eqref{eq:IntToReal}.
Employing the Monte Carlo method, we utilize generated price paths to price a European call option and path-dependent options described in \ref{sec:option}, Asian, lookback, up-and-out barrier call options.
The payoffs in these options, which are originally defined with the entire path in continuous time, are now replaced with the discrete-time versions:
\begin{align}
    f_{\rm pay}(S_{t_1},\cdots,S_{t_M}) = \max \left\{ \frac{1}{M}\sum_{j=1}^M S_{t_j} - K,0 \right\}
\end{align}
for an Asian option,
\begin{align}
    f_{\rm pay}(S_{t_1},\cdots,S_{t_M}) = \max \left\{ \max_{j=1,\cdots,M} S_{t_j} - K , 0 \right\}
\end{align}
for a lookback option, and
\begin{align}
    f_{\rm pay}(S_{t_1},\cdots,S_{t_M}) =
    \begin{cases}
    \max \left\{S_{t_M} - K,0  \right\}& (\max_{j=1,\cdots,M} S_{t_j}  \,  < B  )\\
        0 & (\mathrm{otherwise})
    \end{cases}.
\end{align}
Strikes of these options are set to $K=100$.
As for the up-and-out Barrier option, we set the barrier level $B=105$.
We also compute prices of these options using the paths in the original Heston model generated by Eq.\,\eqref{eq:EulerMaruyama} as a benchmark.
By comparing the results in the two ways, we can analyze the effectiveness and accuracy of the MPS model in pricing these path-dependent options. 

To compare results of the Heston model and the MPS model more closely, we also introduce the {\it implied volatility} (IV) $\sigma_{BS} (K,T)$ as follows:
\begin{align}\label{eq:implied_vol}
    C = C_{BS}(S_0,K,T,\sigma_{BS}(K,T)) 
\end{align}
where $C$ is a price of a European call option with strike $K$ and maturity $T$.
That is, given the market price $C$ we reversely calculate $\sigma_{BS}(K,T)$ from Eq.\,\eqref{eq:implied_vol}.
With $C$ being the market price of the option, we can regard the IV as the market's expectation of the future volatility of the underlying asset.
One reason why the IV is widely used in practice is that it is a strike-independent indicator of the option price level: the option prices for different strikes largely differ, but the IVs are comparable.
It is also common to compare different option pricing methods in terms of IV.

\subsection{Results}

\begin{figure}[t]
    \centering
    \includegraphics[width=1\textwidth]{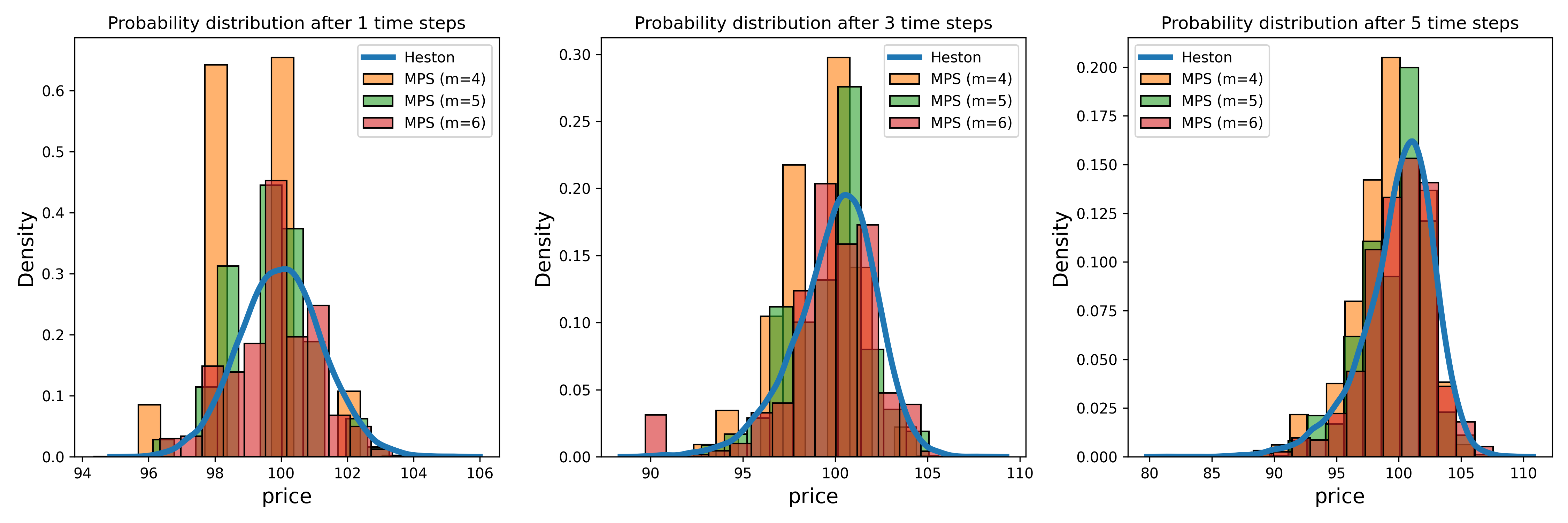}
    \caption{Probability distributions of the asset price at $t=t_1$ (left), $t_3$ (center), and $t_5$ (right) generated by the MPS models with $m=4$ (orange bar), 5 (green bar), and 6 (red bar), and the Heston model (blue curve). $D_{\rm max}$ is fixed to 150. For the Heston model, the Kernel Density Method \cite{silverman2018density} is used to draw continuous lines.}
    \label{fig:distributions}

    \vspace{1em}

    \small
    \begin{tabularx}{\textwidth}{l|ccccc}
             & \multicolumn{2}{c}{European} & Asian  & Lookback & Barrier  \\
             & Price & IV & Price & Price & Price \\
    \hline
    Heston   & 1.1098 (0.0052) & 0.1967  (0.0009) & 0.6195  (0.0035) & 1.4778  (0.0056)   & 0.9894  (0.0049)        \\
    MPS ($m=4$) & 0.6277 (0.0031) & 0.1113   (0.0005) & 0.2771  (0.0020) & 0.8216  (0.0030)  & 0.5769  (0.0028)     \\
    MPS ($m=5$) & 0.8626 (0.0034) & 0.1529   (0.0009) & 0.4351  (0.0021)& 1.1625   (0.0039) & 0.8304  (0.0029)      \\
    MPS ($m=6$) & 1.0805 (0.0041) & 0.1915  (0.0007) & 0.6107   (0.0026) & 1.4612   (0.0049) & 0.9160  (0.0030)      
    \end{tabularx}
    \normalsize
    \captionof{table}{Prices of various options calculated by the Monte Carlo method with paths generated by the Heston model and the MPS models with various physical dimensions. $D_{\rm max}$ is fixed to 150. For a European option, IVs are also shown. We run independent 10 calculations for each combination of option types and path-generation models. The displayed numbers are the averages over 10 runs and the standard errors are also shown in parentheses.}
    \label{tab:result}
\end{figure}

We first consider the effect of physical dimensions $2^m$ in MPS models, which is equal to the number of grid points for the discrete approximation of the original continuous variables $x_j$ and thus determines the resolution of the approximation.
For that purpose, we fix $D_{\rm max} = 150$ and vary the physical dimensions $2^m$ with $m=4,5,6$.
\figref{fig:distributions} shows the resultant probability distributions of $S_{t_1}$, $S_{t_3}$, and $S_{t_5}$ for each physical dimension.
There, we plot empirical distributions of $S_t$ for $N=10000$ samples generated by the Heston and MPS models.
These figures indicate that the higher $m$ becomes the closer the distribution is to the Heston model.
Note that, for $m=4$, which means prices are discretized into only $2^4 = 16$ grid points, the resulting distribution is localized around several values, as in the left graph in \figref{fig:distributions}, but such a phenomenon is mitigated for larger $m$.

We then evaluate the price of options, and the IV calculated from the value of the European call option, showing the result in \tblref{tab:result}.
Similarly to the probability distributions at each time step, with higher physical dimensions, option prices as well as the IV calculated by the MPS model get closer to those by the Heston model. 
On the European option, the IV by the MPS model with $m=6$ has a difference of several tenths of a percent from that by the Heston model, which we can say is practically a good fit for a pricing model of path-dependent options.
Also for the Asian and lookback options,
the prices by the MPS model are close to those by the Heston model with differences similar to that of the European option.
Thus, we can conclude that the MPS model with sufficiently large physical dimension successfully learns the Heston model, at least for the purpose of pricing these options.
On the other hand, for the barrier option, the difference between the prices of the two models is still larger even at $m=6$.
We may improve the MPS model by, e.g., taking larger $m$ or using tensor network architectures other than MPS, which can be considered in future works.

\begin{figure}[t]
    \centering
    \includegraphics[width=1\textwidth]{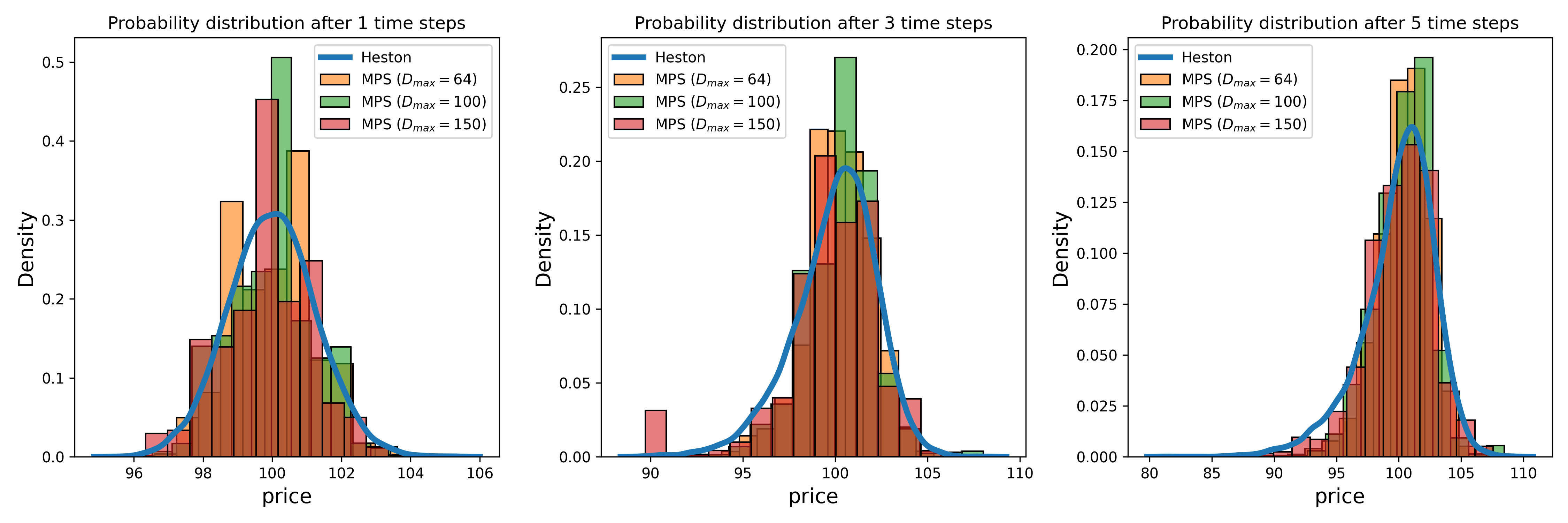}
    \caption{Same as Figure \ref{fig:distributions} except that $D_{\rm max}$ is set to 64 (orange), 100 (green), and 150 (red), with $m$ fixed to 6.}
    \label{fig:distributions_bonddim}
    
    \vspace{1em}
    
    \small
    \tabcolsep=1.5mm
    \begin{tabularx}{\textwidth}{l|ccccc}
             & \multicolumn{2}{c}{European} & Asian  & Lookback & Barrier  \\
             & Price & IV & Price & Price & Price \\
    \hline
    Heston   & 1.1098 (0.0052) & 0.1967  (0.0009) & 0.6195  (0.0035) & 1.4778  (0.0056)  & 0.9894  (0.0049)      \\
    MPS($D_{\rm max}=64$) & 0.9767 (0.0036)  & 0.1731   (0.0006) & 0.5547  (0.0020) & 1.3270  (0.0036)  & 0.9605  (0.0031)    \\
    MPS($D_{\rm max}=100$) & 1.0321 (0.0045) & 0.1829  (0.0008) & 0.5733  (0.0027) & 1.3701  (0.0047) & 0.9468  (0.0026) \\
    MPS($D_{\rm max}=150$) & 1.0805 (0.0041)   & 0.1915  (0.0007) & 0.6107   (0.0026) & 1.4612   (0.0049) & 0.9160  (0.0030)       
    \end{tabularx}
    \normalsize
    \captionof{table}{Same as \tblref{tab:result} expect that the maximum bond dimension $D_{\rm max}$ is varied with $m$ fixed to 6.}
    \label{tab:result_bonddim}
\end{figure}

The MPS model has the other hyperparameter, which would be another factor for the model's capability, that is, the maximal value of the bond dimension $D_{\rm max}$.
In order to evaluate the effect of $D_{\rm max}$, we thus consider the different $D_{\rm max}$ with fixed $m$. In the below, we set $m=6$.
\figref{fig:distributions_bonddim} shows the probability distributions of $S_t$ for various $D_{\rm max}$.
It is observed that with higher $D_{\rm max}$ the distributions are closer to those of the Heston model.
We also investigate the effect of $D_{\rm max}$ in terms of option prices and the IV as shown in \tblref{tab:result_bonddim}, where the results of the Heston and MPS models with $D_{\rm max} = 150$ are same as in \tblref{tab:result}.
We find that, basically, the prices by the MPS model get closer to those by the Heston model as we increase $D_{\rm max}$.
Only for the barrier option, the price difference becomes larger as $D_{\rm max}$ increases, which implies that the MPS model underestimates its value possibly because the MPS model generates more paths that exceed the barrier level, which largely affects the conditional expectation value for the barrier option, than the Heston model.
The outliers in generated price paths can cause the discrepancy as well.
These problems may be resolved by adopting some regularization technique to the model during training and generation, or by using tensor network structures other than MPS, which should be addressed in the future work.

Lastly, let us comment on the gate cost for encoding the MPS into the quantum state by the quantum circuit shown in Sec. \ref{sec:circuit}.
In the setting with $m=6$ and $D_{\rm max}=150$, the resultant MPS has bond dimensions $(D_1,D_2,D_3,D_4)=(16,130,150,32)$, for which the gate cost is of order
\begin{align}
    \sum_{j=1}^{M-1} 4^{d_j+d_{j+1}}+4^{d_M}\simeq 4.8 \times 10^9.
\end{align}
Although, as mentioned in Sec. \ref{sec:circuit}, our method has the gate cost advantage with respect to $M$, the actual gate cost is rather large for the current physical and bond dimensions.
It would be desired to compress the MPS-encoding circuit using some technique such as the ones in \cite{ran2020encoding,rudolph2023decomposition}, which leverages the layer-wise circuit structures.
We will consider such an improvement in future work.

\section{Conclusion}\label{sec:conclusion}
In this paper, we propose to utilize MPS for generating time series. 
Our primary concern is option pricing in quantitative finance, where it is expected that quantum computing can accelerate the speed. 
To this end, we train the MPS model for the Heston model, which is common in the financial industry, and evaluate path-dependent options, using time series samples generated by the trained model.
Our finding is that the proposed MPS model not only successfully generates price paths and probability distributions of the learned Heston model, but also calculates several path-dependent options using these samples with the Monte Carlo method.

We believe that our result sheds light on future implementation of quantum computing in finance, as MPS is naturally embedded into quantum circuits, as shown in Sec. \ref{sec:circuit}, and can prepare quantum states for probability distributions of price paths for option pricing.
Although our MPS-based method has the gate cost advantage with respect to $M$, the number of time points, the actual gate cost is still large for the current physical and bond dimensions, and is desired to be compressed further.

We emphasize that our result differs from existing works in the sense that they focus on generating a probability distribution at some fixed time slice.
Our work enables more flexible generation of time series samples, which gives us many potential applications and research directions.
As a concluding remark, we comment some of them in the following.

While in this study we only consider the Heston model as a model of asset dynamics, there are other stochastic models used in the field of finance, such as the CEV model \cite{cox1996constant}, SABR model \cite{hagan2002managing}, and so on. 
It would be interesting whether the MPS model can also reproduce these models' dynamics.
Another important issue is to implement our model in the quantum circuits, which might be difficult to do in the current environments, but should hopefully be possible in the future and reproduce our result.  
Additionally, it would be meaningful to conduct end-to-end implementation of option pricing in quantum computers with the combination of our MPS model and the Monte Carlo integration in quantum computing.
In this context, along with preparing a state encoding the path distribution, efficient implementations of other potential bottlenecks would be required, e.g., calculating the payoff of a complicated product such as path-dependent ones \cite{Chakrabarti2021thresholdquantum}.

\section*{Acknowledgement}

KM is supported by MEXT Quantum Leap Flagship Program (MEXT Q-LEAP) Grant no. JPMXS0120319794, JSPS KAKENHI Grant no. JP22K11924, and JST COI-NEXT Program Grant No. JPMJPF2014. The authors are grateful to Kosuke Mitarai for helpful discussions.

\appendix

\section{Heston model}\label{sec:Heston}
The Heston model \cite{heston1993closed} is a popular mathematical model to describe the dynamics of the asset price and its volatility and has become widely used in the financial industry.
It is one of the SV models, meaning it incorporates a stochastic process of the volatility as well as the asset price.
SV models can be used when market prices of European options written on the asset do not match the prices given by the BS model, which is an often observed phenomenon called the {\it volatility smile} \cite{dupire1994pricing}.

In the Heston model, the dynamics of the asset is given by the following SDE in the risk-neutral measure:
\begin{align}
    dS_t &= \sqrt{\nu_t} S_t dW_t^S  \label{eq:Heston} \\ 
    d \nu_t &= \kappa(\theta -\nu_t) dt + \xi \sqrt{\nu_t} dW_t^\nu \label{eq:CIR}
\end{align}
where $W_t^S$ and $W_t^\nu$ are one-dimensional Wiener process with correlation $\rho$, which appears as $dW_t^S dW_t^\nu = \rho dt$, and $\kappa, \theta, \xi$ are positive constants.
The Heston model describes the squared volatility $\nu_t$ as another stochastic process in Eq.\,\eqref{eq:CIR}, known as the CIR process \cite{cox2005theory}, which exhibits mean-reverting.
Note that the BS model corresponds to the case where $\nu_t$ is a positive constant, meaning the volatility is not dynamical, as opposed to the Heston model.

\section{Option pricing by quantum Monte Carlo integration \label{sec:QM}}

Here, we briefly review option pricing with a quantum computer.
The process of this algorithm is decomposed into three parts:
\begin{enumerate}
    \item Loading the probability distributions of price paths into a quantum state
    \item Encoding the payoff function of the option into the amplitude of an ancilla qubit
    \item Utilizing QAE to estimate the expectation value of the payoff function
\end{enumerate}
As a first step, we prepare a quantum oracle $\mathcal{P}$, which encodes the probability distribution of price paths into the quantum state in Eq.~\eqref{eq:TargetState}:
\begin{align}
    \mathcal{P} \ket{0} = \sum_{\{S_{t_j}\}_j} \sqrt{p(\{S_{t_j}\}_j)} \ket{\{S_{t_j}\}_j} \, .
\end{align}
Then, in the second step, we operate the oracle $\mathcal{F}$, which acts on an ancilla qubit as 
\begin{align}
    \mathcal{F}\ket{\{S_{t_j}\}_j} \ket{0} = \ket{\{S_{t_j}\}_j}  \left(\sqrt{1-f_{\rm pay}(\{S_{t_j}\}_j)} \ket{0} + \sqrt{f_{\rm pay}(\{S_{t_j}\}_j)} \ket{1}  \right) \, .
\end{align}
As a result, the combination of oracles $\mathcal{Q} = \mathcal{F} \mathcal{P}$ creates a quantum state
\begin{align}
    \sum_{\{S_{t_j}\}_j} \sqrt{p(\{S_{t_j}\}_j)} \ket{\{S_{t_j}\}_j}\left(\sqrt{1-f_{\rm pay}(\{S_{t_j}\}_j)} \ket{0} + \sqrt{f_{\rm pay}(\{S_{t_j}\}_j)} \ket{1}  \right) \, ,
\end{align}
which encodes the desired expectation value in the amplitude, as we see that the probability of measuring $\ket{1}$ in the last qubit is equal to 
\begin{align}
    \sum_{ \{S_{t_j}\}_j}p(\{S_{t_j}\}_j) f_{\rm pay}( \{S_{t_j}\}_j) = E [f_{\rm pay} (\{S_t\})] \, .
\end{align}
We do not cover the last step that uses QAE to elucidate the above probability; interested readers may see \cite{rebentrost2018quantum,stamatopoulos2020option}.
By leveraging QAE, the query complexity to achieve the error $\epsilon$ scales as $O(1/\epsilon)$ while in classical algorithm it increases as $O(1/\epsilon^2)$, which provides the quadratic speed-up of the Monte Carlo method.

\section{Update of tensor components \label{sec:MPSUpdate}}

Here, we explain the procedure of updating tensor components in our MPS training.

Let us suppose that we are updating the $A^{(j)}_n$ and $A^{(j+1)}_n$ to $A^{(j)}_{n+1}$ and $A^{(j+1)}_{n+1}$.
Here and hereafter, the subscript $n\in\{0,1,\ldots\}$ denotes that the tensor is the one after the $n$-th update, with the initial value $A^{(j)}_n$ set randomly from a uniform distribution between 0 and 1.
In the $(n+1)$-th update, we first merge $A^{(j)}_n$ and $A^{(j+1)}_n$ into rank-4 tensors, 
\begin{align}
    \left(A^{(j,j+1)}_n\right)_{\alpha_{j-1} \bar x_{j} \bar x_{j+1} \alpha_{j+1}} = \sum_{\alpha_j} \left(A^{(j)}_n\right)_{\alpha_{j-1} \bar x_{j} \alpha_j} \left(A^{(j+1)}_n\right)_{\alpha_{j} \bar x_{j+1} \alpha_{j+1}} \, . 
\end{align}
Then we compute the gradient of $\mathcal{L}$ with respect to the merged tensor $A^{(j,j+1)}_n$, which is given by 
\begin{align}
    \frac{\partial \mathcal{L}}{\partial \left(A^{(j,j+1)}_n\right)_{\alpha_{j-1} \bar x_{j} \bar x_{j+1} \alpha_{j+1}}} = \frac{1}{|\mathcal{T}|Z}\frac{\partial Z}{\partial \left(A^{(j,j+1)}_n\right)_{\alpha_{j-1} \bar x_{j} \bar x_{j+1} \alpha_{j+1}}} - \frac{2}{|\mathcal{T}|} \sum_{\bar x_t} \frac{1}{\Psi (\bar x_t)}\frac{\partial \Psi (\bar x_t)}{\partial \left(A^{(j,j+1)}_n\right)_{\alpha_{j-1} \bar x_{j} \bar x_{j+1} \alpha_{j+1}}}.
    \label{eq:gradient}
\end{align} 
Using Eq.\,\eqref{eq:gradient}, we update the elements of the merged tensor $A^{(j,j+1)}_n$ as 
\begin{align}
    &\left(A^{(j,j+1)}_n\right)_{\alpha_{j-1} \bar x_{j} \bar x_{j+1} \alpha_{j+1}} \rightarrow \nonumber \\
    & \quad \left(\hat{A}^{(j,j+1)}_n\right)_{\alpha_{j-1} \bar x_{j} \bar x_{j+1} \alpha_{j+1}} \coloneqq \left(A^{(j,j+1)}_n\right)_{\alpha_{j-1} \bar x_{j} \bar x_{j+1} \alpha_{j+1}} - \eta \frac{\partial\mathcal{L}}{\partial \left(A^{(j,j+1)}_n\right)_{\alpha_{j-1} \bar x_{j} \bar x_{j+1} \alpha_{j+1}}}
\end{align}
with a learning rate $\eta$.
After this update, we perform the singular value decomposition for $\hat{A}^{(j,j+1)}_n$:
\begin{align}
    \hat{A}^{(j,j+1)}_n = U \Lambda V^\dagger
\end{align}
Here, $\hat{A}^{(j,j+1)}_n$ is regarded as a $2^mD_{j-1}\times2^mD_{j+1}$ matrix with the index pair $(\alpha_{j-1},\bar x_{j})$ specifying the row and $(\bar x_{j+1},\alpha_{j+1})$ specifying the column.
$U$ and $V$ are orthogonal matrices with size $2^mD_{j-1}\times2^mD_{j-1}$ and $2^mD_{j+1}\times2^mD_{j+1}$, respectively.
$\Lambda$ is a $2^mD_{j-1} \times 2^mD_{j+1}$ diagonal matrix in which the singular values of $\hat{A}^{(j,j+1)}_n$ are arranged in descending order from the $(1,1)$-th to $(D,D)$-th entries, where $D=\min\{2^mD_{j-1},2^mD_{j+1}\}$, and the other entries are 0.
We then truncate the singular values, that is, replace all the entries in $\Lambda$ smaller than $s_1\epsilon_{\rm cutoff}$ with 0, where $s_1$ is the largest singular value, and the truncation threshold $\epsilon_{\rm cutoff}$ is now set to 0.05~\footnote{We saw that varying $\epsilon_{\rm cutoff}$ in the range from $10^{-1}$ to $10^{-4}$ has little effect on the accuracy of our MPS model in the numerical experiment in Sec.~\ref{sec:experiment}.}.
Supposing that $D_j^\prime$ singular values remain, we reach the approximation
\begin{align}
    \hat{A}^{(j,j+1)}_n\approx\tilde{A}^{(j,j+1)}_n \coloneqq \tilde{U} \tilde{\Lambda} \tilde{V}^\dagger,
\end{align}
where $\tilde{\Lambda}$ is the upper-left $D_j^\prime \times D_j^\prime$ block of $\Lambda$, and $\tilde{U}\in\mathbb{R}^{2^mD_{j-1} \times D_j^\prime}$ (resp. $\tilde{V}\in\mathbb{R}^{2^mD_{j+1} \times D_j^\prime}$) consists of the first $D_j^\prime$ columns of $U$ (resp. $V$).
Finally, we update the original rank-3 matrices by
\begin{align}
    A^{(j)}_{n+1}=\tilde{U}\sqrt{\tilde{\Lambda}}\, , \quad A^{(j+1)}_{n+1}=\sqrt{\tilde{\Lambda}}\tilde{V}^\dagger,
\end{align}
where $A^{(j)}_{n+1} \in \mathbb{R}^{D_{j-1} \times 2^m \times D_{j}^\prime}$ is regarded as a $2^mD_{j-1} \times D_j^\prime$ matrix with $(\alpha_{j-1},\bar x_{j})$ specifies the row and $\alpha_{j}$ specifies the column, and $A^{(j+1)}_{n+1} \in \mathbb{R}^{D_{j}^\prime \times 2^m \times D_{j+1}^\prime}$ is regarded as a $D_j^\prime \times D_{j+1}2^m$ matrix with $\alpha_{j}$ specifies the row and $(x_{j+1},\alpha_{j+1})$ specifies the column.
Note that the bond dimension $D_j$ has changed to $D_j^\prime$.

We start the algorithm with performing this procedure for the rightmost pair.
After that, we move on to the next left pair, and repeat the same. 
When reaching the leftmost tensor, we reverse the updating process from left to right.
The whole sweeping loop starting and ending at the rightmost tensor defines one epoch of the training.
We then run some epochs to get the final result.
The bond dimensions $D_j$ are adjusted as explained above, with some criterion for singular value truncation and an upper bound $D_{\rm max}$.
Consult \cite{han2018unsupervised} for further details.

\bibliographystyle{ieeetr}
\bibliography{ref}

\begin{thebibliography}{10}

\bibitem{orus2019quantum}
R.~Or{\'u}s, S.~Mugel, and E.~Lizaso, ``Quantum computing for finance: Overview and prospects,'' {\em Reviews in Physics}, vol.~4, p.~100028, 2019.

\bibitem{bouland2020prospects}
A.~Bouland, W.~van Dam, H.~Joorati, I.~Kerenidis, and A.~Prakash, ``Prospects and challenges of quantum finance,'' {\em arXiv preprint arXiv:2011.06492}, 2020.

\bibitem{egger2020quantum}
D.~J. Egger, C.~Gambella, J.~Marecek, S.~McFaddin, M.~Mevissen, R.~Raymond, A.~Simonetto, S.~Woerner, and E.~Yndurain, ``Quantum computing for finance: State-of-the-art and future prospects,'' {\em IEEE Transactions on Quantum Engineering}, vol.~1, pp.~1--24, 2020.

\bibitem{herman2022survey}
D.~Herman, C.~Googin, X.~Liu, A.~Galda, I.~Safro, Y.~Sun, M.~Pistoia, and Y.~Alexeev, ``A survey of quantum computing for finance,'' {\em arXiv preprint arXiv:2201.02773}, 2022.

\bibitem{herman2023quantum}
D.~Herman, C.~Googin, X.~Liu, Y.~Sun, A.~Galda, I.~Safro, M.~Pistoia, and Y.~Alexeev, ``Quantum computing for finance,'' {\em Nature Reviews Physics}, vol.~5, no.~8, pp.~450--465, 2023.

\bibitem{montanaro2015quantum}
A.~Montanaro, ``Quantum speedup of monte carlo methods,'' {\em Proceedings of the Royal Society A: Mathematical, Physical and Engineering Sciences}, vol.~471, no.~2181, p.~20150301, 2015.

\bibitem{brassard2002quantum}
G.~Brassard, P.~Hoyer, M.~Mosca, and A.~Tapp, ``Quantum amplitude amplification and estimation,'' {\em Contemporary Mathematics}, vol.~305, pp.~53--74, 2002.

\bibitem{rebentrost2018quantum}
P.~Rebentrost, B.~Gupt, and T.~R. Bromley, ``Quantum computational finance: Monte carlo pricing of financial derivatives,'' {\em Physical Review A}, vol.~98, no.~2, p.~022321, 2018.

\bibitem{stamatopoulos2020option}
N.~Stamatopoulos, D.~J. Egger, Y.~Sun, C.~Zoufal, R.~Iten, N.~Shen, and S.~Woerner, ``Option pricing using quantum computers,'' {\em Quantum}, vol.~4, p.~291, 2020.

\bibitem{kaneko2022quantum}
K.~Kaneko, K.~Miyamoto, N.~Takeda, and K.~Yoshino, ``Quantum pricing with a smile: implementation of local volatility model on quantum computer,'' {\em EPJ Quantum Technology}, vol.~9, pp.~1--32, 2022.

\bibitem{grover2002creating}
L.~Grover and T.~Rudolph, ``Creating superpositions that correspond to efficiently integrable probability distributions,'' {\em arXiv preprint quant-ph/0208112}, 2002.

\bibitem{MunozCores2022}
E.~Mu\~{n}oz Coreas and H.~Thapliyal, ``Everything you always wanted to know about quantum circuits,'' {\em Wiley Encyclopedia of Electrical and Electronics Engineering}, pp.~1--17, 2022.

\bibitem{Sanders2019}
Y.~R. Sanders, G.~H. Low, A.~Scherer, and D.~W. Berry, ``Black-box quantum state preparation without arithmetic,'' {\em Phys. Rev. Lett.}, vol.~122, p.~020502, Jan 2019.

\bibitem{wang2021fast}
S.~Wang, Z.~Wang, G.~Cui, S.~Shi, R.~Shang, L.~Fan, W.~Li, Z.~Wei, and Y.~Gu, ``Fast black-box quantum state preparation based on linear combination of unitaries,'' {\em Quantum Information Processing}, vol.~20, no.~8, p.~270, 2021.

\bibitem{Bausch2022fastblackboxquantum}
J.~Bausch, ``Fast {B}lack-{B}ox {Q}uantum {S}tate {P}reparation,'' {\em {Quantum}}, vol.~6, p.~773, Aug. 2022.

\bibitem{Wang_2022}
S.~Wang, Z.~Wang, R.~He, S.~Shi, G.~Cui, R.~Shang, J.~Li, Y.~Li, W.~Li, Z.~Wei, and Y.~Gu, ``Inverse-coefficient black-box quantum state preparation,'' {\em New Journal of Physics}, vol.~24, p.~103004, oct 2022.

\bibitem{mcardle2022quantum}
S.~McArdle, A.~Gily{\'e}n, and M.~Berta, ``Quantum state preparation without coherent arithmetic,'' {\em arXiv preprint arXiv:2210.14892}, 2022.

\bibitem{rattew2022preparing}
A.~G. Rattew and B.~Koczor, ``Preparing arbitrary continuous functions in quantum registers with logarithmic complexity,'' {\em arXiv preprint arXiv:2205.00519}, 2022.

\bibitem{Sanchez2023}
G.~Marin-Sanchez, J.~Gonzalez-Conde, and M.~Sanz, ``Quantum algorithms for approximate function loading,'' {\em Phys. Rev. Res.}, vol.~5, p.~033114, Aug 2023.

\bibitem{Moosa_2023}
M.~Moosa, T.~W. Watts, Y.~Chen, A.~Sarma, and P.~L. McMahon, ``Linear-depth quantum circuits for loading fourier approximations of arbitrary functions,'' {\em Quantum Science and Technology}, vol.~9, p.~015002, oct 2023.

\bibitem{zoufal2019quantum}
C.~Zoufal, A.~Lucchi, and S.~Woerner, ``Quantum generative adversarial networks for learning and loading random distributions,'' {\em npj Quantum Information}, vol.~5, no.~1, p.~103, 2019.

\bibitem{GarciaRipoll2021quantuminspired}
J.~J. Garc{\'{i}}a-Ripoll, ``Quantum-inspired algorithms for multivariate analysis: from interpolation to partial differential equations,'' {\em {Quantum}}, vol.~5, p.~431, Apr. 2021.

\bibitem{Endo2020}
K.~Endo, T.~Nakamura, K.~Fujii, and N.~Yamamoto, ``Quantum self-learning monte carlo and quantum-inspired fourier transform sampler,'' {\em Phys. Rev. Res.}, vol.~2, p.~043442, Dec 2020.

\bibitem{Holmes2020}
A.~Holmes and A.~Y. Matsuura, ``Efficient quantum circuits for accurate state preparation of smooth, differentiable functions,'' in {\em 2020 IEEE International Conference on Quantum Computing and Engineering (QCE)}, pp.~169--179, 2020.

\bibitem{fuchs2023hybrid}
F.~Fuchs and B.~Horvath, ``A hybrid quantum wasserstein gan with applications to option pricing,'' {\em Available at SSRN 4514510}, 2023.

\bibitem{novikov2016exponential}
A.~Novikov, M.~Trofimov, and I.~Oseledets, ``Exponential machines,'' {\em arXiv preprint arXiv:1605.03795}, 2016.

\bibitem{stoudenmire2016supervised}
E.~Stoudenmire and D.~J. Schwab, ``Supervised learning with tensor networks,'' {\em Advances in neural information processing systems}, vol.~29, 2016.

\bibitem{huggins2019towards}
W.~Huggins, P.~Patil, B.~Mitchell, K.~B. Whaley, and E.~M. Stoudenmire, ``Towards quantum machine learning with tensor networks,'' {\em Quantum Science and technology}, vol.~4, no.~2, p.~024001, 2019.

\bibitem{han2018unsupervised}
Z.-Y. Han, J.~Wang, H.~Fan, L.~Wang, and P.~Zhang, ``Unsupervised generative modeling using matrix product states,'' {\em Physical Review X}, vol.~8, no.~3, p.~031012, 2018.

\bibitem{cheng2019tree}
S.~Cheng, L.~Wang, T.~Xiang, and P.~Zhang, ``Tree tensor networks for generative modeling,'' {\em Physical Review B}, vol.~99, no.~15, p.~155131, 2019.

\bibitem{liu2023tensor}
J.~Liu, S.~Li, J.~Zhang, and P.~Zhang, ``Tensor networks for unsupervised machine learning,'' {\em Physical Review E}, vol.~107, no.~1, p.~L012103, 2023.

\bibitem{ran2020encoding}
S.-J. Ran, ``Encoding of matrix product states into quantum circuits of one-and two-qubit gates,'' {\em Physical Review A}, vol.~101, no.~3, p.~032310, 2020.

\bibitem{rudolph2022synergy}
M.~S. Rudolph, J.~Miller, D.~Motlagh, J.~Chen, A.~Acharya, and A.~Perdomo-Ortiz, ``Synergy between quantum circuits and tensor networks: Short-cutting the race to practical quantum advantage,'' {\em arXiv preprint arXiv:2208.13673}, 2022.

\bibitem{rudolph2023decomposition}
M.~S. Rudolph, J.~Chen, J.~Miller, A.~Acharya, and A.~Perdomo-Ortiz, ``Decomposition of matrix product states into shallow quantum circuits,'' {\em Quantum Science and Technology}, vol.~9, no.~1, p.~015012, 2023.

\bibitem{heston1993closed}
S.~L. Heston, ``A closed-form solution for options with stochastic volatility with applications to bond and currency options,'' {\em The review of financial studies}, vol.~6, no.~2, pp.~327--343, 1993.

\bibitem{hull1993options}
J.~Hull, {\em Options, futures, and other derivative securities}, vol.~7.
\newblock Prentice Hall Englewood Cliffs, NJ, 1993.

\bibitem{shreve2005stochastic}
S.~Shreve, {\em Stochastic calculus for finance I: the binomial asset pricing model}.
\newblock Springer Science \& Business Media, 2005.

\bibitem{shreve2004stochastic}
S.~E. Shreve {\em et~al.}, {\em Stochastic calculus for finance II: Continuous-time models}, vol.~11.
\newblock Springer, 2004.

\bibitem{black1973pricing}
F.~Black and M.~Scholes, ``The pricing of options and corporate liabilities,'' {\em Journal of political economy}, vol.~81, no.~3, pp.~637--654, 1973.

\bibitem{Merton1973}
R.~C. Merton, ``Theory of rational option pricing,'' {\em The Bell Journal of Economics and Management Science}, vol.~4, no.~1, pp.~141--183, 1973.

\bibitem{ORUS2014117}
R.~Or\'{u}s, ``A practical introduction to tensor networks: Matrix product states and projected entangled pair states,'' {\em Annals of Physics}, vol.~349, pp.~117--158, 2014.

\bibitem{fishman2013monte}
G.~Fishman, {\em Monte Carlo: concepts, algorithms, and applications}.
\newblock Springer Science \& Business Media, 2013.

\bibitem{miyamoto2023extracting}
K.~Miyamoto and H.~Ueda, ``Extracting a function encoded in amplitudes of a quantum state by tensor network and orthogonal function expansion,'' {\em Quantum Information Processing}, vol.~22, no.~6, p.~239, 2023.

\bibitem{Vartiainen2004}
J.~J. Vartiainen, M.~M\"ott\"onen, and M.~M. Salomaa, ``Efficient decomposition of quantum gates,'' {\em Phys. Rev. Lett.}, vol.~92, p.~177902, Apr 2004.

\bibitem{Mottonen2005}
M.~M\"{o}tt\"{o}nen, J.~J. Vartiainen, V.~Bergholm, and M.~M. Salomaa, ``Transformation of quantum states using uniformly controlled rotations,'' {\em Quantum Info. Comput.}, vol.~5, p.~467–473, Sept. 2005.

\bibitem{Maruyama1955}
G.~Maruyama, ``Continuous markov processes and stochastic equations,'' {\em Rend. Circ. Mat. Palermo}, vol.~4, pp.~48--90, 1955.

\bibitem{silverman2018density}
B.~W. Silverman, {\em Density estimation for statistics and data analysis}.
\newblock Routledge, 2018.

\bibitem{cox1996constant}
J.~C. Cox, ``The constant elasticity of variance option pricing model,'' {\em Journal of Portfolio Management}, p.~15, 1996.

\bibitem{hagan2002managing}
P.~S. Hagan, D.~Kumar, A.~S. Lesniewski, and D.~E. Woodward, ``Managing smile risk,'' {\em Wilmott}, vol.~1, pp.~84–--108, 2002.

\bibitem{Chakrabarti2021thresholdquantum}
S.~Chakrabarti, R.~Krishnakumar, G.~Mazzola, N.~Stamatopoulos, S.~Woerner, and W.~J. Zeng, ``A {T}hreshold for {Q}uantum {A}dvantage in {D}erivative {P}ricing,'' {\em {Quantum}}, vol.~5, p.~463, June 2021.

\bibitem{dupire1994pricing}
B.~Dupire {\em et~al.}, ``Pricing with a smile,'' {\em Risk}, vol.~7, no.~1, pp.~18--20, 1994.

\bibitem{cox2005theory}
J.~C. Cox, J.~E. Ingersoll~Jr, and S.~A. Ross, ``A theory of the term structure of interest rates,'' in {\em Theory of valuation}, pp.~129--164, World Scientific, 2005.

\end{thebibliography}

\end{document}